\documentclass[twocolumn,twoside,slac]{revtex4}
\usepackage{graphicx}
\usepackage{fancyhdr}
\begin{document}

\title{The New Object Oriented Analysis Framework For H1}

%

\author{M. Peez on behalf of the H1 Collaboration}
\affiliation{CPPM, IN2P3-CNRS, Marseille, France}

\begin{abstract}

During the years 2000 and 2001 the HERA machine and the H1 experiment 
performed substantial luminosity upgrades. To cope with the increased demands
on data handling an effort was made to redesign and modernize the analysis 
software. Main goals were to lower turn-around time for physics analysis
by providing a single framework for data storage, event selection, physics and
event display.
The new object oriented analysis environment based on the RooT framework
provides a data access front-end for the new data storage scheme and
a new event display. The analysis data is stored in four different
layers of separate files. Each layer represents a different level of
abstraction, i.e. reconstruction output, physics particles, event summary
information and user specific information. Links between the layers allow 
correlating quantities of different layers. 
Currently, this framework is used for data analyses of the previous collected
data and for standard data production of the currently collected data.

\end{abstract}

\maketitle

\thispagestyle{fancy}


\section{Introduction}
H1 is an experiment at the electron proton collider HERA at DESY
(Germany), which started collecting data in 1992.
Essential parts of the analysis software architecture as well
as the basic data model were established ten years ago.
Like most of the high energy physics experiments at that time
H1 used FORTRAN based packages such as 
BOS \cite{BOS} and FPACK for data storage and access. 
Physics analysis were performed on so-called n-tuples
using HBOOK \cite{HBOOK} and PAW \cite{PAW}.
After the HERA and H1 upgrade in the year 2000, the increased
luminosity put new demands on data storage and data handling. 
Therefore, the H1 Collaboration has decided to move towards a new technology
and to develop an analysis environment which should:
\begin{itemize}
\item lower turn around time for physics analysis
\item provide a unique, modern, extendable and re-usable framework
\item incorporate and support all H1 physics analyses
\item standardize the physics algorithms, e.g.~kinematic reconstruction,
selection criteria, particle identifications etc.
\item provide one unique code reference and therefore
facilitate exchange of information between different analysis groups
\item make expert knowledge reusable by non-experts and
lower the threshold of starting a new analysis
\item provide a faster, more efficient access to the data
\item make doing analyses in H1 more attractive to students
\end{itemize}
To cope with these requirements, the H1 Collaboration
chose to base its analysis software on the object oriented analysis
framework RooT \cite{ROOT}. RooT is based on C++ and provides software for 
all aspects related to physics analysis 
(i.e. processing, storing and visualizing data).
In addition a C++ interpreter allows interactive use of the framework within
the same scripting language.\\

\section{The data storage model}
In order to standardize physics analysis and to take full advantage of the new 
partial event reading capability of the RooT framework a four layer data
 structure has been implemented.
\begin{itemize}
\item {\bf ODS ($Object~Data~Store$):}
This layer is a 1-1 copy of the reconstruction output formerly called
'Data Summary Tape' (DST). 
It contains reconstructed tracks, clusters as well as important
detector informations. Even though this layer can be stored
persistently, the standard is to produce it transiently from DST.
This way people doing analysis in the old and those using the new framework
can work in parallel without having to store the same information twice.
The size of ODS objects is of the order of 13 kb/event.
\item {\bf $\mu$ODS ($\mu Object~Data~Store$):}
On $\mu$ODS 4-vectors of physical particles and particle candidates
are stored (see fig. \ref{muODS}). 
The sum of all particle candidates provides a 4 $\Pi$ coverage
of the solid angle around the reconstructed vertex and there
is no double counting of energy. Each of the particles stores also
a pointer to the ODS-tracks and clusters that it was build from.\\
The identified, physical particles contain all particle information
and some specific detector informations that was used to identify the particle
or that might be necessary for further specification during physics
analysis. Composed particles (as for example jets or J/Psi particles)
are stored in a special class containing pointers to the related particle
candidates or identified particles. With the provided information, the
$\mu$ODS is largely sufficient for most analysis purposes. A mean amount
of 3 kb/event has been achieved.
\begin{figure*}[t]
\centering
\includegraphics[width=100mm, bb = 1 1 519 371]{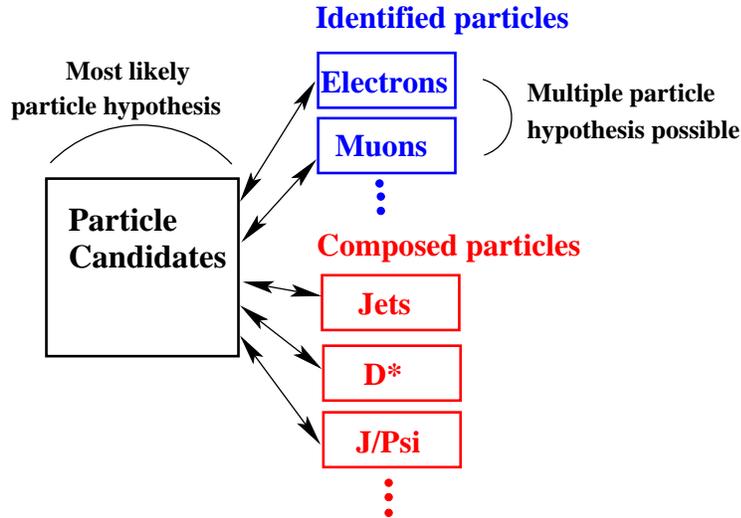}
\caption{Structure of the $\mu$ODS level with pointers between the
identified and composed particles.} \label{muODS}
\end{figure*}
\item {\bf HAT:}
The HAT ('H1 Analysis Tag') contains event summary level information, such as 
event kinematics and particle multiplicities.
All variables  are stored as basic types.
The main purpose it to allow a fast event selection (``tagging'') for
each of the physics analysis.
The size is 0.4 kb/event. 
\item {\bf UserTree:} 
To persistenly store user specific information
that is not already stored on official data layers a so-called
``user-tree'' is supported by the framework.
\end{itemize}

The ODS layer is filled transiently 
using DST information only. The $\mu$ODS is filled
by the physics algorithm detailed in the section \ref{physics_algo}.
New $\mu$ODS and HAT are centrally produced whenever new analysis algorithms or calibration results are available.

\section{The data access}
The data access is implemented in a set of skeleton classes
which were developed by following the three main requirements:
\begin{itemize}
\item the user has only one single event loop (synchronization of the different
layers is transparent to the user)
\item a transparent access to the different levels of the data: 'Smart'
accessor functions allow to retrieve information about event and
particle attributes across boundaries of different files, e.g. $\mu$ODS-to-ODS.
\item the access to the data is partial, e.g. accessing a cluster on ODS from
$\mu$ODS should not require to read the full ODS event.
\end{itemize}
These requirements are implemented in the class 'H1Tree' and some helper
classes by using different RooT trees in parallel, one for each storage level.
One of these helper classes is 'H1EventList' which is based on the
RooT TEventList and
facilitates the access to events according to user selection.
H1EventList allows to cumulate different selections and to select
data on each layer.

\section{The physics algorithm \label{physics_algo}}
As H1 is a running experiment with an increasing flow of new data, it 
is essential that the quality and precision of these physics analysis be 
sustained. Therefore the first goal was, while learning from the already
existing algorithms in FORTRAN, to develop and implement algorithms in the
new framework with better performances than the old ones.
To ensure quality and extendibility of the new analysis
software, a modular organization
of loosely physics algorithms is essential. 
The aim is to allow
for routines developed in particular user analyses to
be integrated in the official production code and in addition
to facilitate the physics analysis and to lower the turn-around time
for beginners.
 Modularity and portability is a
prerequisite for the goal that the best knowledge of all physics working 
groups in H1 be propagated into one common framework.
The interface between the filling code and the physics algorithms is 
structured such that the addition of new algorithms involves minimal changes
to the software. Technically the algorithms
are implemented in separate classes that obey the same interface.
The running is divided into two steps:
\begin{itemize}
\item First, the particle finders reconstruct 
the identified particles and the particle candidates
using ODS objects only.
\item In a second step, the composed particle identifiers run on the
already reconstructed particles.
\end{itemize}
The first category of finders comprises an electron, muon, hadron and photon
finder. They are based on the already existing algorithms in fortran
and show the same performances in terms of quality and precision. 
In the second category of finders, a jet finder using a $K_{T}$ algorithm
as well as a J/$\Psi$, a $D^{*}$ and K$_{0}$ have been implemented
and validated.
In future, new particle finder could easely be integrated in the new scheme
and the existings ones are continuously be improved.

\section{The event display}
A new event display has been developed. It is an application
based on the new physics analysis framework
and thus allows for the direct dialogue between the
analysis part (e.g. event selection and histogramming) and visual inspection
of events. The display was originally derived from the
Alice 3D RooT display. Thanks to the RooT Run Time Information (RTTI),
objects on the screen can be picked and inspected, thus accessing the
physics information. Graphics sliders can be used to 
apply selections such that 
only relevant objects for a certain analysis are displayed.
A new feature is the possibility to
display the particle 4-vectors stored on $\mu$ODS on top
of the detector objects.
For instance, one could display the 4-vectors of particles (different
particle types are displayed in different colours)
on top of the detector information and require minimum transverse momentum
for all particles by moving a graphic slider.
One advantage of the new display is his full backward compatibility to the
old 2D command-line base program based on LOOK \cite{LOOK}. Existing
code containing expert knowledge about detector details is reused, thus
fully integrating the functionality of the previous display. A new
parser for the 'LOOK'-macro language was written in C++. 
It is possible to display event information stored in RooT files as well
as information stored in the former DST format. Raw information, like
for example the hit information of a reconstructed track, could
therefore easely be retrieved.  
The new
program combines modern features, such as the GUI, the click and inspect 
options and the 3D-display with the advantages of the old display.

\section{Summary}
A new analysis framework based on object oriented programming techniques
has been succesfully introduced. The key to this success was
the clear definition of the scope of the project:
\begin{itemize}
\item A code development group takes care of the technical challenges, such
as encapsulation of the data handling. 
The physics working groups develop algorithms and add their code
via well defined interfaces. The main reconstruction algorithms have
been implemented, tested and validated.
\item The end users obtain a nice and easy-to-use product integrating all 
analysis specific tools into one single framework.
\end{itemize}
The physics analysis all greatly profit from the new and enhanced analysis 
environment. The framework is widely accepted within the Collaboration:  almost
all of the new starting analyses are based on the framework. It is in addition
 used for the official data quality checks.


\end{document}